\documentstyle[12pt]{article}


\newlength{\minitwocolumn}
\catcode`\@=11
\long\def\@makefntext#1{
\protect\noindent \hbox to 3.2pt {\hskip-.9pt  
$^{{\eightrm\@thefnmark}}$\hfil}#1\hfill}               

\def\@makefnmark{\hbox to 0pt{$^{\@thefnmark}$\hss}}    
        
\def\ps@myheadings{\let\@mkboth\@gobbletwo
\def\@oddhead{\hbox{}
\rightmark\hfil\eightrm\thepage}   
\def\@oddfoot{}\def\@evenhead{\eightrm\thepage\hfil
\leftmark\hbox{}}\def\@evenfoot{}
\def\sectionmark##1{}\def\subsectionmark##1{}}

\font\eightrm=cmr8

\font\sc=cmr5 scaled\magstep1

\def\PL{Phys.~Lett. }

\def\brs{{\delta_{\hbox{\sc B}}}}

\def\delnl{\delta_{\hbox{\sc NL}}}
\def\ep{\epsilon}

\def\delzero{\delta_0}
\def\brs{\delta}

\def\bold#1{#1\llap{$#1$\hskip.3pt}\llap{$#1$\hskip.4pt}\llap{$#1$\hskip.5pt}}
\def\bbold#1{\bold{\bold#1}}

\def\bos#1{\bbold#1}

\def\ba{{\bos{A}}}
\def\bb{{\bos{B}}}

\def\bphi{{\bos{\phi}}}

\def\bDelta{{\bos{\Delta}}}
\def\bdelta{{\bos{\delta}}}
\newcommand{\dbv}{{\Delta_{\hbox{\sc BV}}}}
\newcommand{\OmBV}{{\Omega_{\hbox{\sc BV}}}}
\newcommand{\bOmBV}{\bos{\Omega_{\hbox{\sc BV}}}}

\begin{document}



\pagestyle{empty}

\begin{center}
{\large\bf 
Deformation of BF theories,
Topological Open Membrane and 
A Generalization of The Star Deformation 
}

\vspace{15mm}

Noriaki IKEDA
\footnote{ E-mail address:\ ikeda@yukawa.kyoto-u.ac.jp} \\
Ritsumeikan University\\
Kusatsu, Shiga 525-8577, Japan \\
and \\
Setsunan University \\
Neyagawa, Osaka 572-8508, Japan 
\end{center}
\date{}


\vspace{15mm}
\begin{abstract}
We consider a deformation of the BF theory in any dimension by means
of the antifield BRST formalism. 
Possible consistent interaction terms for the action and the gauge
symmetries are analyzed and we find a new class of topological gauge
theories.
Deformations of the world volume BF theory are considered as 
possible deformations of the topological open membrane.
Therefore if we consider these theories on open membranes,
we obtain noncommutative structures of the boundaries of open
membranes, and we propose
a generalization of the path integral representation of the star
deformation.  
\end{abstract}

\newpage
\pagestyle{plain}
\pagenumbering{arabic}


\rm
\section{Introduction}
\noindent
Kontsevich has given
a general formula for the deformation quantization \cite{Bayen:1978ha}
of functions on the Poisson manifold in the paper \cite{Ko}.
Cattaneo and Felder \cite{CF} have obtained the path integral
representation of that formula on the Poisson manifold as a
perturbative expansion of a two-dimensional field theory on 
two-dimensional disk.
The star product structure in the open string theory 
with non-zero background Neveu-Schwarz B-field
appears essentially at the same mechanism \cite{SW}.

The world sheet field theory in the paper \cite{CF} 
is called a nonlinear gauge theory or the Poisson sigma model in 
two-dimension 
\cite{II1}\cite{SS}. 
It is one of Schwarz type (or BF type) topological field
theory \cite{BBRT} and has the gauge symmetry, which is a
generalization of the usual nonabelian gauge symmetry.

Izawa \cite{Iz} has recently analyzed the nonlinear gauge 
theory from the viewpoint of 
a deformation of the gauge symmetry \cite{BH}.
He has found that two-dimensional nonlinear gauge theory is the unique 
consistent deformation of two-dimensional abelian BF
theory.
%
The author has made the similar analysis in three dimension \cite{I2}
\footnote{A deformation of the BF theories 
has analyzed in a special case by Dayi \cite{Day}.}.
He has considered the abelian BF theory in three dimension and all their
possible deformations which action has the ghost number zero,
and has found the new topological gauge theory.
In this paper, we make a similar analysis in any dimension.
We consider the abelian BF theory and analyze all deformations by the 
antifield BRST formalism.

We consider the two-dimensional theory as a survey of our story.
First we consider the following action of the abelian BF 
theory in two dimension:
\begin{eqnarray}
S_0 = \int {\cal L}_0, 
\qquad 
{\cal L}_0 = B_a d \phi^a,
\label{twoabelian}
\end{eqnarray}
where $a, b$, etc. are Lie algebra indices (or the target space indices).
$\phi^a$ is a scalar field, $B_a$ is a one-form gauge field.
This theory has the following abelian gauge symmetry:
\begin{eqnarray}
\delzero B_a = d \ep_a,
\qquad
\delzero \phi^a = 0.
\label{ags}
\end{eqnarray}
Of course, the commutation relation on the scalar field $\phi$ is 
$[ \phi^a, \phi^b ] = 0$ and correlation functions of arbitrary functions 
of $\phi$ are simply products of one-point functions.

Here we find all the consistent interactions with the free action 
(\ref{ags}) up to field redefinition as a local field theory in two
dimension with the aid of Barnich-Henneaux method 
in terms of the BRST cohomology 
based on the antifield (Batalin-Vilkovisky) formalism.
We use the BRST formalism to deform the world sheet theory, in order
to realize the ``physical requirement''.
In the papers \cite{BH}, they have required 
locality, unitarity and gauge invariance of the deformed action 
and have given a method to analyze the BRST cohomology and obtained
all deformations of the gauge theory.

The answer in two dimension is 
\begin{eqnarray}
S = \int {\cal L}, 
\qquad 
{\cal L} = B_a d \phi^a + \frac{1}{2} W^{ab} B_a B_b,
\label{twodim}
\end{eqnarray}
under the assumption that the ghost number of the action is zero, where
$W^{ab}(\phi) = - W^{ba}(\phi)$ is an arbitrary function of $\phi^a$
\cite{Iz}.
The gauge symmetry is deformed to the nonlinear gauge symmetry:
\begin{eqnarray}
\delnl B_a = d \ep_a + \frac{\partial W^{bc}} {\partial \phi^a} 
B_b \ep_c,
\qquad
\delnl \phi^a = W^{ba} \ep_b,
\label{nlgs}
\end{eqnarray}
where $W^{ab}(\phi)$ must satisfy the following identities:
\begin{eqnarray}
\frac{\partial W^{ab}} {\partial \phi^d} W^{cd}
+ \frac{\partial W^{bc}} {\partial \phi^d} W^{ad}
+ \frac{\partial W^{ca}} {\partial \phi^d} W^{bd}= 0,  
\label{WJacobi}
\end{eqnarray}
in order for (\ref{nlgs}) to be a symmetry of the theory.
This Eq.(\ref{WJacobi}) is just the Jacobi identity if the following
commutation relation holds: 
\begin{eqnarray}
[ \phi^a, \phi^b ] = W^{ab}(\phi).
\label{comm}
\end{eqnarray}
%
The commutation relation on the left hand side 
is realized as the Poisson bracket of the coordinates $\phi^a$ and
$\phi^b$ on the Poisson manifold \cite{SS}. 
If we consider the theory on the two-dimensional disk,
the correlation functions at the boundary of the disk of arbitrary
functions of $\phi$ are deformed to the star product $F*G$ 
under the appropriate regularization and the appropriate boundary
condition \cite{CF}.

Recently, noncommutative geometry are widely used to analyze
the string theory. 
If we consider the open string theory with the constant background NS
B-field, the noncommutative geometry appears on the D-brane \cite{SW}.
The generalization of the star product to the nonconstant B-field has
been anlayzed in \cite{Cornalba:2001sm}.
Some authors have analyzed generalizations of this theory to the
higher dimension.
Noncommutative geometry appears at the boundary of open 2-brane in
M-theory \cite{BBSS}\cite{KS}\cite{Das:2001mg}.
In this paper, we propose one approach to analyze them.
We consider that deformations of the world volume BF theory lead us to
deformations of the boundaries of the topological open $n-1$-brane,
where $n-1$ is the space dimension of the membrane.
In fact, the star deformation formula is derived from the deformation of
two-dimensional BF theory as the world sheet theory. 

Deformations of the topological open string have been analyzed in 
\cite{Hofman:2001ce}.
Deformations of the topological open membrane in three dimension
have been analyzed in \cite{Park} \cite{Hofman:2001zt}.
We will obtain a generalization of
the path integral representation of the star deformation formula. 
However analysis of deformations of the BF theories are not still
completed. In this paper, we analyze deformations of the BF theory.

This paper is organized as follows.
In section 2, we construct the superfield antifield formalism
of the abelian BF theory.
In section 3, we analyze deformations of the abelian BF theory 
and obtain all possible deformations.
In section 4, we consider two examples of our theory in the lower
dimensions.
In section 5, we consider the quantum BV formalism and quantize the theory.
In section 6, we analyze the topological membrane action.
Section 7 is conclusion and discussion.

\newcommand\gh{{\rm gh}}
\newcommand{\lb}[2]{[\![#1\,,#2]\!]}
\newcommand{\rd}{\overleftarrow{\partial}} 
\newcommand{\ld}{\overrightarrow{\partial}} 
\newcommand{\sbv}[2]{{\left(\!\left({\,{#1}\,,\,{#2}\,}\right)\!\right)}}

\section{The Superfield Formalism for the Batalin-Vilkovisky Action of
the Abelian BF Theory}
\noindent
First, we consider the following $n$-dimensional abelian BF theory:
\begin{eqnarray}
S_0 = \sum_{p=0}^{[\frac{n-1}{2}]} \int_{\Sigma} 
(-1)^{n-p} B_{n-p-1\ a} d A_p{}^a,
\label{abf}
\end{eqnarray}
where 
$A_p{}^a$ is a p-form gauge field and 
$B_{n-p-1\ a}$ is a $n-p-1$ form auxiliary fields.
Indices $a, b, c$, etc. represent algebra indices.
$\Sigma$ is a base manifold on which the theory is defined.
The sign factors $(-1)^{n-p}$ are introduced for convenience.
%
%
%
%
This action has the following abelian gauge symmetry:
\begin{eqnarray}
&& \delzero A_{p}{}^{a} = d c_{p-1}^{(p)a}, \nonumber \\
&& \delzero B_{n-p-1 \ a} 
= d t_{n-p-2\ a}^{(n-p-1)},
\label{abrs}
\end{eqnarray}
where $c_{p-1}^{(p)a}$ is a $p-1$-form gauge parameter and $t_{n-p-2\
a}^{(n-p-1)}$ is a $n-p-2$-form gauge parameter.
$(p)$ in $c_{p-1}^{(p)a}$ and $(n-p-1)$ in $t_{n-p-2\ a}^{(n-p-1)}$ 
represent that $c_{p-1}^{(p)a}$ is a gauge
parameter for $p$-form $A_{p}{}^{a}$
and $t_{n-p-2\ a}^{(n-p-1)}$ is one for $n-p-1$-form $B_{n-p-1 \ a}$,
respectively.
%
This gauge symmetry is reducible.
Since $A_{p}{}^{a}$ is a $p$-form and 
$B_{n-p-1 \ a}$ is a $n-p-1$-form,
we need the following towers of the 'ghost for ghosts' to analyze the
complete gauge degrees of freedom:
\begin{eqnarray}
&& \delzero A_{p}{}^{a} = d c_{p-1}^{(p)a}, 
\qquad 
\delzero B_{n-p-1 \ a} 
= d t_{n-p-2\ a}^{(n-p-1)}, \nonumber \\ 
&& \delzero c_{p-1}^{(p)a} = d c_{p-2}^{(p)a},
\qquad
\delzero t_{n-p-2\ a}^{(n-p-1)} 
= d t_{n-p-3\ a}^{(n-p-1)}, \nonumber \\
&& \vdots  \nonumber \\
&& \delzero c_{1}^{(p)a} = d c_0^{(p)a}, 
\qquad
\delzero t_{1\ a}^{(n-p-1)} =  d t_{0\ a}^{(n-p-1)},
\nonumber \\
&& \delzero c_0^{(p)a} = 0, 
\qquad
\delzero t_{0\ a}^{(n-p-1)} = 0,
\label{agauge}
\end{eqnarray}
where $c_{i}^{(p)a}$ are $i$-form gauge parameters and $t_{j\
a}^{(n-p-1)}$ are $j$-form gauge parameters.
$i = 0, \cdots, p-1$ and $j = 0, \cdots, n-p-2$.

We write the theory
by the antifield BRST formalism.
First we take $c_{i}^{(p)a}$ to be the
FP ghosts $i$-form with ghost number $p-i$, and $t_{j\ a}^{(n-p-1)}$ 
to be a $j$-form with the ghost number $n-p-1-j$.
As usual, if the ghost number is odd,
the fields are Grassmann odd, and if ghost number even,
they are Grassmann even.

Next we introduce the antifields for all the fields.
Let $\Phi^+$ denote the antifields for the field $\Phi$.
Note that the relations ${\rm deg}(\Phi) + {\rm deg}(\Phi^+) = n$ and
${\rm gh}(\Phi) + {\rm gh}(\Phi^+) = -1$ are required,
where ${\rm deg}(\Phi)$ and ${\rm deg}(\Phi^+)$ are the form degrees 
of the fields $\Phi$ and $\Phi^+$
and ${\rm gh}(\Phi)$ and ${\rm gh}(\Phi^+)$ are the ghost numbers of
them.
For $A_{p}{}^{a}$, we introduce the antifield $A_{n-p\ a}^{+(p)}$
, which is $n-p$-form with the ghost number $-1$.
For $B_{n-p-1 \ a}$, $B_{p+1}^{+(n-p-1)a}$, which is  
$p+1$-form with the ghost number $-1$.
For $c_{i}^{(p)a}$, $c_{n-i \ a}^{+(p)}$, which is  
$n-i$-form with the ghost number $-p-1+i$.
For $t_{j\ a}^{(n-p-1)}$, $t_{n-j}^{+(n-p-1)a}$, which is  
$n-j$-form with the ghost number $-n+p+j$.

For functions $F(\Phi, \Phi^+)$ and $G(\Phi, \Phi^+)$ of the fields
and the antifields,
we define the antibracket as follows;
\begin{eqnarray}
(F, G) \equiv \frac{F \rd}{\partial \Phi} \frac{\ld G}{\partial \Phi^+}
- (-1)^{(n+1) \deg \Phi}
\frac{F \rd}{\partial \Phi^+} \frac{\ld G}{\partial \Phi},
\label{anti}
\end{eqnarray}
where ${\rd}/{\partial \varphi}$ and ${\ld}/{\partial
\varphi}$ are the right differentiation and the left differentiation
with respect to $\varphi$, respectively.
The following identity about left and right derivative is useful:
\begin{eqnarray}
\frac{\ld F}{\partial \varphi} =  (-1)^{(\gh F - \gh \varphi) \gh \varphi 
+ (\deg F - \deg \varphi) \deg \varphi}
\frac{F \rd}{\partial \varphi}.
\label{lrdif}
\end{eqnarray}
If $S, T$ are two functionals, the antibracket is defined as follows:
\begin{eqnarray}
(S, T) \equiv \int_{\Sigma}
\left(
\frac{S \rd}{\partial \Phi} \frac{\ld T}{\partial \Phi^+}
- (-1)^{(n+1) \deg \Phi}
\frac{S \rd}{\partial \Phi^+} \frac{\ld T}{\partial \Phi}.
\right)
\label{antif}
\end{eqnarray}
The antibracket satisfies the following identities:
\begin{eqnarray}
&& (F, G) = -(-1)^{(\deg F - n)(\deg G - n) + (\gh F + 1)(\gh G + 1)}(G, F),
\nonumber \\
&& (F, GH) = (F, G)H + (-1)^{(\deg F -n)\deg G + (\gh F +1) \gh G} G(F, H),
\nonumber \\
&& (FG, H) = F(G, H) + (-1)^{\deg G(\deg H -n) + \gh G(\gh H +1) } (F, H)G,
\nonumber \\
&& (-1)^{(\deg F - n)(\deg H - n) + (\gh F + 1)(\gh H + 1) } (F, (G, H)) 
+ {\rm cyclic \ permutations} = 0,
\label{antibra}
\end{eqnarray}
where $F, G$ and $H$ are functions on fields and antifields.

In order to simplify 
notations and calculations, we rewrite notations by the
superfield formalism.
%
We combine the field, its antifield and their gauge descendant fields
as superfield components. 
%
For $A_{p}{}^{a}$ and $B_{n-p-1 \ a}$, 
we define corresponding superfields as follows:
\begin{eqnarray}
\ba_p{}^a &=& 
c_0^{(p)a} + c_{1}^{(p)a} 
+ \cdots 
+ c_{p-1}^{(p)a} 
+ A_{p}{}^{a} 
+ B_{p+1}^{+(n-p-1)a} 
\nonumber \\ && 
+ t_{p+2}^{+(n-p-1)a} 
+ \cdots
+ t_{n}^{+(n-p-1)a}, 
\nonumber \\
%
\bb_{n-p-1\ a} &= &
t_{0\ a}^{(n-p-1)} 
+ t_{1 \ a}^{(n-p-1)}
+ \cdots 
+ t_{n-p-2 \ a}^{(n-p-1)} 
+ B_{n-p-1 \ a} 
+ A_{n-p \ a}^{+(p)} 
\nonumber \\ && 
+ c_{n-p+1 \ a}^{+(p)} 
+ \cdots
+ c_{n \ a}^{+(p)}.
\label{component}
\end{eqnarray}
Then we define the total degree $|F| \equiv \gh F + \deg F$.
The component fields in a superfield have the same total degree.
The total degrees of 
$\ba_p{}^a$ and $\bb_{n-p-1\ a}$ are $p$ and $n-p-1$, respectively.

We take a notation $\cdot$ as
the {\it dot product} among superfields in order to simplify the sign 
factors \cite{Cattaneo:2000mc}.
The definitions and properties of the {\it dot product} are in the appendix.
The {\it dot antibracket} of the superfields $F$ and $G$ is defined as
\begin{eqnarray}
\sbv{F}{G} \equiv (-1)^{(\gh F + 1) (\deg G - n)} 
(-1)^{\gh \Phi (\deg \Phi - n) + n} 
(F, G), 
\label{dotantibra}
\end{eqnarray}
Then the following identities are obtained from the equations
(\ref{antibra}) and (\ref{dotantibra}):
\begin{eqnarray}
&& \sbv{F}{G} = -(-1)^{(|F| + 1 - n)(|G| + 1 - n)} \sbv{G}{F},
\nonumber \\
&& \sbv{F}{GH} = \sbv{F}{G} \cdot H 
+ (-1)^{(|F| + 1 - n)|G|} G \cdot \sbv{F}{H},
\nonumber \\
&& \sbv{FG}{H} = F \cdot \sbv{G}{H} 
+ (-1)^{|G|(|H| + 1 - n)} \sbv{F}{H} \cdot G,
\nonumber \\
&& (-1)^{(|F| + 1 - n)(|H| + 1 - n)} \sbv{F}{\sbv{G}{H}} 
+ {\rm cyclic \ permutations} = 0.
\end{eqnarray}
We define the {\it dot differential} as
\begin{eqnarray}
&& \frac{\ld }{\partial \varphi} \cdot F 
\equiv (-1)^{\gh \varphi \deg F}
\frac{\ld F}{\partial \varphi}, \nonumber \\
&&F \cdot \frac{\rd }{\partial \varphi}
\equiv (-1)^{\gh F \deg \varphi}
\frac{F \rd}{\partial \varphi}.
\end{eqnarray}
Then, ~from the equation (\ref{lrdif}), we can obtain the formula
\begin{eqnarray}
\frac{\ld }{\partial \varphi} \cdot F =   
(-1)^{(|F| - |\varphi|) |\varphi|}
F \cdot \frac{\rd }{\partial \varphi}.
\end{eqnarray}
Since $\ba_p^a$ and $\bb_{n-p-1 \ a}$ are the field-antifield pair,
we can rewrite the BV antibracket on two superfields $F$ and $G$ 
from (\ref{anti}) and (\ref{dotantibra}) as follows:
\begin{eqnarray}
\sbv{F}{G} \equiv 
\sum_{p=0}^{[\frac{n-1}{2}]} 
F \cdot \frac{\rd}{\partial \ba_p{}^a} \cdot
\frac{\ld }{\partial \bb_{n-p-1 \ a}} \cdot G
- (-1)^{n p}
F \cdot \frac{\rd }{\partial \bb_{n-p-1 \ a}} \cdot
\frac{\ld }{\partial \ba_p{}^a} \cdot G.
\end{eqnarray}
%

We can construct the Batalin-Vilkovisky action for the abelian BF theory
by the superfields as follows:
\begin{eqnarray}
S_0 = \sum_{p=0}^{[\frac{n-1}{2}]} \int_{\Sigma}
(-1)^{n-p} \bb_{n-p-1\ a} \cdot d \ba_p{}^a,
\label{sabf}
\end{eqnarray}
where we integrate only $n$-form part of the integrand.
If we set all the antifields zero,
(\ref{sabf}) reduces to 
(\ref{abf}).
The BRST transformation for the superfield $F$ under the
action above is defined as
\begin{eqnarray}
\delzero F = \sbv{S_0}{F}=
\sum_{p=0}^{[\frac{n-1}{2}]} S_0 \cdot \frac{\rd }{\partial \ba_p{}^a} \cdot
\frac{\ld }{\partial \bb_{n-p-1 \ a}} \cdot F
- (-1)^{n p}
S_0 \cdot \frac{\rd }{\partial \bb_{n-p-1 \ a}} \cdot
\frac{\ld }{\partial \ba_p{}^a} \cdot F,
\end{eqnarray}
Hence the BRST transformations on $\ba_p{}^a$ and $\bb_{n-p-1\ a}$ are
obtained as follows:
\begin{eqnarray}
\delzero \ba_p{}^a
&=& \sbv{S_0}{\ba_p{}^a}=
(-1)^{n-p} \frac{\ld }{\partial \bb_{n-p-1\ a}} \cdot S_0 \nonumber \\
&=&
d \ba_p{}^a, \nonumber \\
\delzero \bb_{n-p-1\ a}
&=& \sbv{S_0}{\bb_{n-p-1 \ a}}=
(-1)^{p(n-p)} \frac{\ld }{\partial \ba_p{}^a} \cdot S_0 \nonumber \\
&=&
d \bb_{n-p-1\ a}.
\label{ababelBRST}
\end{eqnarray}
If we expand the BRST transformations (\ref{ababelBRST}) above 
to the components by (\ref{component}), 
we can obtain the BRST transformation on each field and antifield
as follows:
\begin{eqnarray}
&& \delzero c_0^{(p)a} = 0, \nonumber \\
&& \delzero c_{1}^{(p)a} = d c_0^{(p)a}, \nonumber \\ 
&& \vdots  \nonumber \\
&& \delzero c_{p-1}^{(p)a} = d c_{p-2}^{(p)a}, \nonumber \\ 
&& \delzero A_{p}{}^{a} = d c_{p-1}^{(p)a}, \nonumber \\
&& \delzero B_{p+1}^{+(n-p-1)a} = d A_{p}{}^{a}, \nonumber \\
&& \delzero t_{p+2}^{+(n-p-1)a} = d B_{p+1}^{+(n-p-1)a}, \nonumber \\ 
&& \delzero t_{p+3}^{+(n-p-1)a} = d t_{p+2}^{+(n-p-1)a}, \nonumber \\
&& \vdots  \nonumber \\
&& \delzero t_{n}^{+(n-p-1)a} = d t_{n-1}^{+(n-p-1)a}, 
\nonumber \\
%
&& \delzero t_{0\ a}^{(n-p-1)} = 0, \nonumber \\
&& \delzero t_{1\ a}^{(n-p-1)} = d t_{0\ a}^{(n-p-1)},
\nonumber \\
&& \vdots \nonumber \\
&& \delzero t_{n-p-2\ a}^{(n-p-1)} 
= d t_{n-p-3\ a}^{(n-p-1)}, \nonumber \\
&& \delzero B_{n-p-1 \ a} 
= d t_{n-p-2\ a}^{(n-p-1)}, \nonumber \\ 
&& \delzero A_{n-p \ a}^{+(p)} = d B_{n-p-1 \ a}, 
\nonumber \\
&& \delzero c_{n-p+1 \ a}^{+(p)} = 
d A_{n-p \ a}^{+(p)}, \nonumber \\
&& \delzero c_{n-p+2 \ a}^{+(p)} 
= d c_{n-p+1 \ a}^{+(p)}, \nonumber \\ 
&& \vdots \nonumber \\
&& \delzero c_{n \ a}^{+(p)} = d c_{n-1 \ a}^{+(p)},
\end{eqnarray}
which reproduce the gauge transformations (\ref{agauge}).

$S_0$ must be BRST invariant. In fact, 
\begin{eqnarray}
\delzero S_0 = \sbv{S_0}{S_0}
&=& 2 \sum_{p=0}^{[\frac{n-1}{2}]} (-1)^{(n-p)} \int_{\Sigma}
d (\bb_{n-p-1\ a} \cdot d \ba_p{}^a) \nonumber \\
&=& 2 \sum_{p=0}^{[\frac{n-1}{2}]} \int_{\Sigma}
d (d \bb_{n-p-1\ a} \cdot \ba_p{}^a),
\label{zerobrs}
\end{eqnarray}
therefore if the base manifold $\Sigma$ has no boundary, 
$\delzero S_0 = 0$.
If $\Sigma$ has a boundary 
(For example, the open string or the open membrane.),
we can take two kinds of boundary conditions 
${\ba_p{}^a}_{//}|_{\partial \Sigma} = 0$ or 
${\bb_{n-p-1\ a}}_{//}|_{\partial \Sigma} = 0$, 
where the notation ${//}$ mean the components 
along the direction tangent to the boundary $\partial \Sigma$.
We can also take the different boundary condition on each field
component so as to satisfy BRST invariant condition
of the action.
Two boundary conditions are used to construct 
A, B and C boundary conditions in the paper \cite{Park}.
In the rest of this paper, we select appropriate boundary conditions
so as to satisfy $\delzero S_0 = 0$.

It is simple to confirm $\delzero^2=0$ on all superfields.
Equations of motion are
\begin{eqnarray}
&& d \ba_p{}^a = 0, 
\qquad 
d \bb_{n-p-1\ a} = 0.
\label{aeqom}
\end{eqnarray}
\section{Deformation of BF Theory}
\noindent
Let us consider a deformation of the action $S_0$
perturbatively,
\begin{eqnarray}
&& S = S_0 + g S_1 + g^2 S_2 + \cdots,
\label{pertur}
\end{eqnarray}
where $g$ is a deformation parameter, or a coupling constant of the
theory. The total BRST transformation is deformed to
\begin{eqnarray}
&&\brs \ba_p{}^a = \sbv{S}{\ba_p{}^a} =
(-1)^{n-p} \frac{\ld }{\partial \bb_{n-p-1\ a}} \cdot S, \nonumber \\
&&\brs \bb_{n-p-1\ a} = \sbv{S}{\bb_{n-p-1\ a}}
= (-1)^{p(n-p)} \frac{\ld}{\partial \ba_{p}{}^a} \cdot S.
\end{eqnarray}
In order for the deformed BRST transformation $\brs$ to be nilpotent
and make the theory consistent,
the total action $S$ has to satisfy the following classical master
equation:
\begin{eqnarray}
\sbv{S}{S} = 0.
\label{master}
\end{eqnarray}
Substituting (\ref{pertur}) to (\ref{master}), we obtain 
\begin{eqnarray}
\sbv{S}{S} = 
\sbv{S_0}{S_0} 
+ 2g \sbv{S_0}{S_1}
+ g^2 [ \sbv{S_1}{S_1} + 2 \sbv{S_0}{S_2} ] + O(g^3) = 0.
\label{purmaster}
\end{eqnarray}
We solve this equation order by order.
At the $0$-th order, we obtain $\delzero S_0 = \sbv{S_0}{S_0} =0$, which
is already satisfied from (\ref{zerobrs}).
At the first order of $g$ in the Eq.~(\ref{purmaster}), 
\begin{eqnarray}
\delzero S_1 = \sbv{S_0}{S_1} =0,
\label{1brst}
\end{eqnarray}
is required. 
Now we assume that $S_1$ is given by a {\it local} Lagrangian: 
\begin{eqnarray}
&&
S_1 = \int_{\Sigma} {\cal L}_1,
\label{s1act} 
\end{eqnarray}
where ${\cal L}_1$ is constructed from the superfields 
$\ba_p{}^a$ and $\bb_{n-p-1\ a}$ with
$p=0, \cdots, {[\frac{n-1}{2}]}$.
If a monomial in ${\cal L}_1$ includes a differentiation $d$, 
its term is proportional to the equations of motion (\ref{aeqom}).
Therefore its term can be absorbed to the abelian BF theory action
(\ref{sabf}) by the local field redefinitions of $\ba_p{}^a$ or
$\bb_{n-p-1\ a}$, and that term is BRST trivial at the BRST
cohomology\cite{BH}.
%
Hence the nontrivial deformation terms must not include
the differentiation $d$ and 
we can write the candidate ${\cal L}_1$ as 
\begin{eqnarray}
&& S_1 = \int_{\Sigma} {\cal L}_1, \nonumber \\
&& {\cal L}_1 
= \sum_{p_1, \cdots, p_k, q_1, \cdots, q_l}
F_{p_1 \cdots p_k, q_1 \cdots q_l \ a_1 \cdots a_k}
{}^{b_1 \cdots b_l}(\ba_0{}^a)
\cdot \ba_{p_1}{}^{a_1} \cdots \ba_{p_k}{}^{a_k}
\cdot \bb_{q_1 b_1} \cdots \bb_{q_l b_l},
\label{s1}
\end{eqnarray}
where $F_{p_1 \cdots p_k, q_1 \cdots q_l \ a_1 \cdots a_k} 
{}^{b_1 \cdots b_l}(\ba_0{}^a)$ is a function of $\ba_0{}^a$,
and $p_r \neq 0, q_s \neq 0$ for $r = 1, \cdots, k, s = 1, \cdots, l$.
In order to consider the general deformations,
we do not require the total degree of ${\cal L}_1$ is $n$.
Then (\ref{1brst}) is calculated as follows:
\begin{eqnarray}
\delzero S_1 &=& 
\sum_{p_1, \cdots, p_k, q_1, \cdots, q_l} \int_{\Sigma}
[d F_{p_1 \cdots p_k, q_1 \cdots q_l \ a_1 \cdots a_k} 
{}^{b_1 \cdots b_l}(\ba_0{}^a)
\cdot \ba_{p_1}{}^{a_1} \cdots \ba_{p_k}{}^{a_k}
\cdot \bb_{q_1 b_1} \cdots \bb_{q_l b_l} \nonumber \\
&& + 
\sum_{r=1}^k (-1)^{p_1 + \cdots + p_{r-1}}
F_{p_1 \cdots p_k, q_1 \cdots q_l \ a_1 \cdots a_k} 
{}^{b_1 \cdots b_l}(\ba_0{}^a)
\cdot \ba_{p_1}{}^{a_1} \cdots d \ba_{p_r}{}^{a_r} \cdots \ba_{p_k}{}^{a_k}
\cdot \bb_{q_1 b_1} \cdots \bb_{q_l b_l} \nonumber \\
&& + 
\sum_{s=1}^l (-1)^{p_1 + \cdots + p_k + q_1 + \cdots + q_{s-1}}
\nonumber \\
&& \times F_{p_1 \cdots p_k, q_1 \cdots q_l \ a_1 \cdots a_k} 
{}^{b_1 \cdots b_l}(\ba_0{}^a)
\cdot \ba_{p_1}{}^{a_1} \cdots \ba_{p_k}{}^{a_k}
\cdot \bb_{q_1 b_1} \cdots d \bb_{q_s b_s} \cdots \bb_{q_l b_l}]
\nonumber \\
%
&=& 
\sum_{p_1, \cdots, p_k, q_1, \cdots, q_l} \int_{\Sigma}
d [F_{p_1 \cdots p_k, q_1 \cdots q_l \ a_1 \cdots a_k} 
{}^{b_1 \cdots b_l}(\ba_0{}^a)
\cdot \ba_{p_1}{}^{a_1} \cdots \ba_{p_k}{}^{a_k}
\cdot \bb_{q_1 b_1} \cdots \bb_{q_l b_l}],
\end{eqnarray}
and 
$\delzero S_1 = 0$ if 
\begin{eqnarray}
(F_{p_1 \cdots p_k, q_1 \cdots q_l \ a_1 \cdots a_k} 
{}^{b_1 \cdots b_l}(\ba_0{}^a)
\cdot \ba_{p_1}{}^{a_1} \cdots \ba_{p_k}{}^{a_k}
\cdot \bb_{q_1 b_1} \cdots \bb_{q_l b_l})_{//}|_{\partial \Sigma} = 0.
\label{s1boundary}
\end{eqnarray}
$S_1$ must be constructed from the terms which satisfy the
requirements above.
If there is no boundary, there is no restriction for $S_1$.
If we take the boundary condition 
${\ba_p{}^a}_{//}|_{\partial \Sigma} = 0$,
then (\ref{s1boundary}) is satisfied if 
the terms include at least one ${\ba_p{}^a}$. 
If we take ${\bb_{n-p-1\ a}}_{//}|_{\partial \Sigma} = 0$, 
then (\ref{s1boundary}) is satisfied if 
the terms include at least one $\bb_{n-p-1\ a}$.

At the second order of $g$, 
\begin{eqnarray}
\sbv{S_1}{S_1} + 2 \sbv{S_0}{S_2} = 0, 
\label{pursec}
\end{eqnarray}
is required.
We cannot construct nontrivial $S_2$ to satisfy (\ref{pursec})
from the integration of a local Lagrangian, because
$\delzero$-BRST transforms of the local terms are always total
derivative.
Therefore if we assume locality of the action, $S_2$ is a BRST
trivial and we can set $S_i = 0$ for $i \geq 2$.
%
Then the condition (\ref{pursec}) reduces to
\begin{eqnarray}
\sbv{S_1}{S_1} = 0.
\label{s1s1}
\end{eqnarray}
This imposes the identities on the structure functions 
$F_{p_1 \cdots p_k q_1 \cdots q_l \ a_1 \cdots a_k} {}^{b_1 \cdots
b_l}(\ba_0{}^a)$ in (\ref{s1}).
Now we have obtained the possible deformations of the BF theory in any
dimension as
\begin{eqnarray}
S = S_0 + g S_1,
\label{lapS}
\end{eqnarray}
where $S_0$ is (\ref{sabf}) and $S_1$ is defined as (\ref{s1}).
In the next section, we consider two nontrivial examples.
General algebra structure underline the antifield BRST formalism
is the $L_{\infty}$-algebra (the sh Lie algebra) \cite{LS}\cite{Sta}
which is derived from the analysis of the master equation.
The gauge symmetry in our theory generally has extended 
structures of usual Lie algebra, and 
if a deformation satisfies the master equation (\ref{master}),
that is, (\ref{s1s1}), the symmetry of the deformed theory generally
has $L_{\infty}$-algebra structure.

The total BRST transformations $\brs$ for the superfields are as follows:
\begin{eqnarray}
\brs \ba_p{}^a
&=&  (-1)^{n-p} \sbv{S}{\ba_p{}^a} \nonumber \\
&=&
d \ba_p{}^a
+ (-1)^{n-p} \frac{\ld }{\partial \bb_{n-p-1\ a}} \cdot S_1, \nonumber \\
\brs \bb_{n-p-1\ a}
&=& (-1)^{p(n-p)} 
\sbv{S}{\bb_{n-p-1\ a}} \nonumber \\
&=& 
d \bb_{n-p-1\ a}
+ (-1)^{p(n-p)} \frac{\ld}{\partial \ba_{p}{}^a} \cdot S_1.
\label{totalBRST}
\end{eqnarray}
Equations of motion are obtained as
\begin{eqnarray}
&& d \ba_p{}^a
+ (-1)^{n-p} \frac{\ld }{\partial \bb_{n-p-1\ a}} \cdot S_1 = 0, \nonumber \\
&& 
d \bb_{n-p-1\ a}
+ (-1)^{p(n-p)} \frac{\ld}{\partial \ba_{p}{}^a} \cdot S_1 = 0,
\end{eqnarray}
If we set all antifields $\Phi^+ =0$ in (\ref{pertur}), 
we obtain the usual classical action. 
\section{Examples in The Lower Dimensions}
\noindent
In this section, we consider some nontrivial examples of the
deformations in the lower dimensions.
For simplicity, we only consider the action which total ghost number 
is zero.

In two dimension, (\ref{pertur}) is written as
\begin{eqnarray}
&& S = S_0 +g S_1,  \nonumber \\
&& S_0 = \int_{\Sigma}
\bb_{1 a} \cdot d \bphi^a, 
\qquad
S_1 = \int_{\Sigma} \frac{1}{2} f^{ab}(\bphi^a)
\cdot \bb_{1 a} \cdot \bb_{1 b},
\label{2daction}
\end{eqnarray}
where 
$\bphi^a = \ba_0{}^a$ and $\frac{1}{2}f^{ab}(\bphi^a) = F_{,11}(\ba_0{}^a)$.
The condition (\ref{s1s1}) 
imposes the following identity on $f^{ab}$: 
\begin{eqnarray}
\frac{\partial f^{ab}} {\partial \bphi^d} f^{cd}
+ \frac{\partial f^{bc}} {\partial \bphi^d} f^{ad}
+ \frac{\partial f^{ca}} {\partial \bphi^d} f^{bd}= 0.
\label{2dJacobi}
\end{eqnarray}
We find that this theory is nothing but two-dimensional
nonlinear gauge theory (the Poisson sigma model)\cite{II1}\cite{SS}.
We have deformed the abelian BF theory $S_0$ to $S$
by the analysis of the BRST cohomology.
If we relax the condition that the action has ghost number zero,
we can generalize the theory to the polyvector fields, and we obtain
$L_{\infty}$-algebra by the deformation of the abelian BF theory.

Next, we consider the theory in three dimension.
We restrict $S_1$ to the ghost number zero for simplicity again. 
Then the total action (\ref{pertur}) is deformed as follows:
\begin{eqnarray}
&& S = S_0 +g S_1,  \nonumber \\
&& S_0 = \int_{\Sigma}
[- \bb_{2 a} \cdot d \bphi{}^a + \bb_{1 a} \cdot d \ba_1{}^a], 
\nonumber \\
&& S_1 = \int_{\Sigma} d^3 \theta d^3 x
[f_1{}_a{}^b(\bphi) \cdot \ba_1{}^a \cdot \bb_{2 b} 
+ f_2^{ab}(\bphi) \cdot \bb_{2 a} \cdot \bb_{1 b}
+ \frac{1}{3!} f_{3abc}(\bphi) \cdot \ba_1{}^a \cdot \ba_1{}^b 
\cdot \ba_1{}^c
\nonumber \\
&& 
+ \frac{1}{2} f_{4ab}{}^c(\bphi) \cdot \ba_1{}^a \cdot \ba_1{}^b 
\cdot \bb_{1 c}
+ \frac{1}{2} f_{5a}{}^{bc}(\bphi) \cdot \ba_1{}^a \cdot \bb_{1 b}
\cdot \bb_{1 c}
+ \frac{1}{3!} f_6{}^{abc}(\bphi) \cdot \bb_{1 a} \cdot \bb_{1 b}
\cdot \bb_{1 c}],
\label{3daction}
\end{eqnarray}
where we replace the notations as
$f_1{}_a{}^b = F_{1,2 a}{}^b$,
$f_2^{ab} = F_{,21}{}^{ab}$,
$\frac{1}{3!}f_{3abc} = F_{111, abc}$,
$\frac{1}{2}f_{4ab}{}^c = F_{11,1 ab}{}^c$,
$\frac{1}{2}f_{5a}{}^{bc} = F_{1,11 a}{}^{bc}$,
$\frac{1}{3!}f_6{}^{abc} = F_{,111 }{}^{abc}$,
for clarity. 
The condition of the classical
master equation (\ref{s1s1}) imposes the following identities on six
$f_i$'s, $i=1, \cdots, 6$: 
\begin{eqnarray}
&& 
f_{1}{}_e{}^a f_{2}{}^{be} + f_{2}{}^{ae} f_{1}{}_e{}^b = 0, 
\label{jac1} \\
&& 
\frac{\partial f_{1}{}_c{}^a}{\partial \phi^e} f_{1}{}_b{}^e
- \frac{\partial f_{1}{}_b{}^a}{\partial \phi^e} f_1{}_c {}^e
+ f_1{}_e{}^a f_{4bc}{}^e + f_{2}{}^{ae} f_{3ebc} = 0, 
\label{jac2} \\
&&
- f_1{}_b{}^e \frac{\partial f_{2}{}^{ac}}{\partial \phi^e} 
+ f_{2}{}^{ec} \frac{\partial f_1{}_b{}^a}{\partial \phi^e} 
+ f_1{}_e{}^a f_{5b}{}^{ec} - f_{2}{}^{ae} f_{4eb}{}^c = 0, 
\label{jac3} \\
&&
f_{2}{}^{eb} \frac{\partial f_{2}{}^{ac}}{\partial \phi^e} 
- f_{2}{}^{ec} \frac{\partial f_{2}{}^{ab}}{\partial \phi^e} 
+ f_1{}_e{}^a f_{6}^{ebc} + f_{2}{}^{ae} f_{5e}{}^{bc} = 0, 
\label{jac4} \\
&&
f_1{}_{[a}{}^e \frac{\partial f_{4bc]}{}^d}{\partial \phi^e} 
- f_{2}{}^{ed} \frac{\partial f_{3abc}}{\partial \phi^e} 
+ f_{4e[a}{}^d f_{4bc]}{}^{e} + f_{3e[ab} f_{5c]}{}^{de} = 0, 
\label{jac5} \\
&&
f_1{}_{[a}{}^e \frac{\partial f_{5b]}{}^{cd}}{\partial \phi^e} 
+ f_{2}{}^{e[c} \frac{\partial f_{4ab}{}^{d]}}{\partial \phi^e} 
+ f_{3eab} f_6{}^{ecd} 
+ f_{4e[a}{}^{[d} f_{5b]}{}^{c]e} + f_{4ab}{}^e f_{5e}{}^{cd} = 0, 
\label{jac6} \\
&&
f_1{}_a{}^e \frac{\partial f_{6}{}^{bcd}}{\partial \phi^e} 
- f_{2}{}^{e[b} \frac{\partial f_{5a}{}^{cd]}}{\partial \phi^e} 
+ f_{4ea}{}^{[b} f_6{}^{cd]e} + f_{5e}{}^{[bc} f_{5a}{}^{d]e} = 0, 
\label{jac7} \\
&&
f_{2}{}^{e[a} \frac{\partial f_{6}{}^{bcd]}}{\partial \phi^e} 
+ f_6{}^{e[ab} f_{5e}{}^{cd]} = 0, 
\label{jac8} \\
&&
f_1{}_{[a}{}^e \frac{\partial f_{3bcd]}}{\partial \phi^e} 
+ f_{4[ab}{}^{e} f_{3cd]e} = 0,
\label{jac9}
\end{eqnarray}
where $[\cdots]$ on the indices represents the antisymmetrization for
the indices.
For example, $\Phi_{[ab]} = \Phi_{ab} - \Phi_{ba}$.
We find that the action (\ref{3daction}) with
(\ref{jac1})-(\ref{jac9}) is the same theory constructed in the paper
\cite{I2}, 
and a nontrivial consistent deformation of the three-dimensional BF
theory. 
If $\Sigma$ has boundaries, possible terms are restricted according to 
the boundary conditions, which are discussed in section 3.

\section{Quantum BV formalism}
\subsection{Quantum Master Equation}
\noindent
In this section, we consider the quantum theory.
We introduce the Hodge dual fields of the antifields 
$\Phi^{*a} = * \Phi^{+a}$.
We define the BV Laplacian as follows:
\begin{eqnarray}
\dbv F \equiv \sum_{a}
(-1)^{\gh \Phi_a}\frac{\delta}{\delta \Phi_a}
\frac{\delta}{\delta \Phi^{*a}} F,
\label{bvlaplacian}
\end{eqnarray}
where $\frac{\delta}{\delta \Phi_a}$ and 
$\frac{\delta}{\delta \Phi^{*a}}$ are differentiations with respect to 
coefficient functions of the forms.
Then the following identity is satisfied:
\begin{eqnarray}
\dbv (F G) = 
(\dbv F) G + (-1)^{\gh F + n \deg F} dv (F, G) 
+ (-1)^{\gh F} F \dbv G,
\label{bviden}
\end{eqnarray}
where $dv$ is the volume form on $\Sigma$.

In order for the generating functional to be gauge invariant, 
The following quantum master equation is required: 
\begin{eqnarray}
(S, S) - 2 i \hbar \dbv S = 0,
\label{qmast}
\end{eqnarray}
for the quantum action $S$.
${\cal O}$ is an observable if an operator ${\cal O}$ satisfies the
following equation:
\begin{eqnarray}
(S, {\cal O}) - i \hbar \dbv {\cal O}  = 0.
\label{obsbrs}
\end{eqnarray}
If we define an operator $\OmBV$ as
$\OmBV \equiv S - i \hbar \dbv,$
then (\ref{obsbrs}) is denoted as 
$\OmBV {\cal O} = 0.$
%
In our BF theory, we can confirm $\dbv S = 0$, therefore the quantum
master equation (\ref{qmast}) becomes
\begin{eqnarray}
(S, S) = 0.
\label{qmast2}
\end{eqnarray}

Let us rewrite the above formulae on the superfields.
We define the {\it dot } BV Laplacian on the superfield as follows:
\begin{eqnarray}
\bDelta F \equiv (-1)^{\deg F -n} \dbv F,
\end{eqnarray}
Then identity (\ref{bviden}) is rewritten as 
\begin{eqnarray}
\bDelta(F \cdot G) = 
(\bDelta F) \cdot G + (-1)^{(n+1)|F|} dv \sbv{F}{G} 
+ (-1)^{|F|} F \cdot \bDelta G,
\end{eqnarray}
The quantum master equation (\ref{qmast}) is rewritten as 
\begin{eqnarray}
\sbv{S}{S} - 2 i \hbar \bDelta S = 0,
\end{eqnarray}
and the observable condition (\ref{obsbrs}) of an operator ${\cal O}$
becomes 
\begin{eqnarray}
\sbv{S}{{\cal O}} - i \hbar \bDelta {\cal O}  = 0.
\label{obsbrssuper}
\end{eqnarray}
We define that
$\bOmBV = \bdelta - i \hbar \bDelta$,
then (\ref{obsbrssuper}) is written as 
$\bOmBV {\cal O} = 0$.
%
(\ref{qmast2}) becomes that $\sbv{S}{S} =0$.

\subsection{Gauge Fixing}
\noindent
In order to quantize the gauge theory, we must fix the gauge.
We introduce the new fields; Fadeev-Popov antighosts and 
Lagrange multiplier fields (Nakanishi-Lautrup fields). 
Since $\ba_p{}^a$ has the $p$-th order reducible gauge transformation, 
the following many antighosts and NL-fields are needed
\cite{Batalin:1983jr}\cite{Gomis:1995he}.
We introduce antighost and NL-field
pairs, 
$\bar{c}^{k(p)}_{p-1-L\ a}$ and 
$\bar{b}^{k(p)}_{p-1-L\ a}$ 
where $k=0, 2, \cdots, 2 \left[\frac{L}{2}\right]$, and
$c^{k(p)a}_{p-1-L}$ and 
$b^{k(p)a}_{p-1-L}$ 
where $k=1, 3, \cdots, 
2 \left[\frac{L-1}{2}\right]+1$
at the $L$-th reducible order, where $L = 0, 1, \cdots, p-1$.
The four kinds of fields are all $p-1-L$-forms.
$\bar{c}^{k(p)}_{p-1-L\ a}$ has the ghost number $k-L-1$, 
$\bar{b}^{k(p)}_{p-1-L\ a}$ has the ghost number $k-L$, 
$c^{k(p)a}_{p-1-L}$ has the ghost number $L-k$, 
and
$b^{k(p)a}_{p-1-L}$ has the ghost number $L-k+1$.
Moreover we introduce the antifields for the fields above as
$\bar{c}^{+k(p)a}_{n-p+1+L}$ and $\bar{b}^{+k(p)a}_{n-p+1+L}$,
where $k=0, 2, \cdots, 2 \left[\frac{L}{2}\right]$,
$c^{+k(p)}_{n-p+1+L\ a}$ and $b^{+k(p)}_{n-p+1+L\ a}$,
where $k=1, 3, \cdots, 2 \left[\frac{L-1}{2}\right]+1$.
The form degrees and the ghost numbers of the antifields are
determined from the relations ${\rm deg}(\Phi) + {\rm deg}(\Phi^+) =
n$ and ${\rm gh}(\Phi) + {\rm gh}(\Phi^+) = -1$.

For $\bb_{n-p-1\ a}$, we also introduce antighost and NL-field
pairs as $\bar{t}^{k(n-p-1)a}_{n-p-2-L}$ and
$\bar{s}^{k(n-p-1)a}_{n-p-2-L}$, where $k=0, 2, \cdots, 
2 \left[\frac{L}{2}\right]$, and
$t^{k(n-p-1)}_{n-p-2-L\ a}$ and $s^{k(n-p-1)}_{n-p-2-L\ a}$,
where $k=1, 3, \cdots, 2 \left[\frac{L-1}{2}\right]+1$
at the $L$-th reducible order, where $L = 0, 1, \cdots, n-p-2$.
The four kinds of fields are all $n-p-2-L$-forms.
$\bar{t}^{k(n-p-1)a}_{n-p-2-L}$ has the ghost number $k-L-1$, 
$\bar{s}^{k(n-p-1)a}_{n-p-2-L}$ has the ghost number $k-L$,
$t^{k(n-p-1)}_{n-p-2-L\ a}$ has the ghost number $L-k$,
and $s^{k(n-p-1)}_{n-p-2-L\ a}$ has the ghost number $L-k+1$,
We introduce the antifields for the fields above, 
$\bar{t}^{+k(n-p-1)}_{p+2+L\ a}$ and $s^{+k(n-p-1)a}_{p+2+L\
a}$, 
where $k=0, 2, \cdots, 2 \left[\frac{L}{2}\right]$,
and $t^{+k(n-p-1)a}_{p+2+L}$ and $\bar{s}^{+k(n-p-1)a}_{p+2+L}$,
where $k=1, 3, \cdots, 2 \left[\frac{L-1}{2}\right]+1$.

We define the extended BV action including the antighosts and
NL-fields as $S + S_{{\hbox{\sc aux}}}$, where  
\begin{eqnarray}
S_{{\hbox{\sc aux}}}
&& =
(-1)^n \int \sum_{p=1}^{[\frac{n-1}{2}]} \biggl(
\sum_{{k=0 \atop k \ {\rm even}}}^{p-1} 
\sum_{L=k}^{p-1}
(-1)^{|\bar{c}^{k(p)}_{p-1-L\ a}|}
\bar{c}^{+k(p)a}_{n-p+1+L} \bar{b}^{k(p)}_{p-1-L\ a}
\nonumber \\ &&
+
\sum_{{k=1 \atop k \ {\rm odd}}}^{p-1} 
\sum_{L=k}^{p-1}
(-1)^{|c^{k(p)a}_{p-1-L}|}
c^{+k(p)}_{n-p+1+L\ a} b^{k(p)a}_{p-1-L}
\biggr) \nonumber \\
&&+
(-1)^n \int \sum_{p=0}^{[\frac{n-1}{2}]} \biggl(
\sum_{{k=0 \atop k \ {\rm even}}}^{n-p-2} 
\sum_{L=k}^{n-p-2}
(-1)^{|\bar{t}^{k(n-p-1)a}_{n-p-2-L}|}
\bar{t}^{+k(n-p-1)}_{p+2+L\ a} \bar{s}^{k(n-p-1)a}_{n-p-2-L}
\nonumber \\ &&
+
\sum_{{k=0 \atop k \ {\rm odd}}}^{n-p-2} 
\sum_{L=k}^{n-p-2}
(-1)^{|t^{k(n-p-1)}_{n-p-2-L\ a}|}
t^{+k(n-p-1)a}_{p+2+L} s^{k(n-p-1)}_{n-p-2-L\ a}
\biggr).
\end{eqnarray}
Then the BRST transformations of the fields are calculated from the
antibrackets as follows: 
\begin{eqnarray}
&& \brs \bar{c}^{k(p)}_{p-1-L\ a} = \bar{b}^{k(p)}_{p-1-L\ a},
\qquad \brs \bar{b}^{k(p)}_{p-1-L\ a} =0, \nonumber \\ 
&& \brs c^{k(p)a}_{p-1-L} = b^{k(p)a}_{p-1-L}, \qquad 
\brs b^{k(p)a}_{p-1-L} = 0, \nonumber \\
&& \brs \bar{t}^{k(n-p-1)a}_{n-p-2-L} = \bar{s}^{k(n-p-1)a}_{n-p-2-L},
\qquad
\brs \bar{s}^{k(n-p-1)a}_{n-p-2-L} = 0, \nonumber \\
&& \brs t^{k(n-p-1)}_{n-p-2-L\ a} = s^{k(n-p-1)a}_{n-p-2-L\ a},
\qquad \brs s^{k(n-p-1)}_{n-p-2-L\ a} = 0.
\end{eqnarray}
We take the boundary condition as a constant at the boundary for
antighosts, and as zero at the boundary for NL-fields.
%
%
%
Now we fix the gauge as follows:
\begin{eqnarray}
&& S_{{\hbox{\sc GF}}} = S + S_{{\hbox{\sc aux}}}|_{\Phi^+ 
= \frac{\ld \Psi}{\partial \Phi}},
\end{eqnarray}
where $\Psi$ is a gauge fixing fermion which is a function of the
fields $\Phi$ with ghost number $-1$.
For example, if we take the Landau gauge,
the gauge fixing fermion is
\begin{eqnarray}
\Psi
&=& 
\sum_{p=1}^{[\frac{n-1}{2}]} 
\int 
\bar{c}^{(p)}_{p-1\ a} d * A_p{}^a
+ \sum_{L=1}^{p-1} \sum_{{k=0 \atop k \ {\rm even}}}^{L} 
\bar{c}^{k(p)}_{p-1-L\ a} d * c_{p-L}^{k-1(p)a}
+ \sum_{L=1}^{p-1} \sum_{{k=1 \atop k \ {\rm odd}}}^{L} 
d * \bar{c}^{k-1(p)}_{p-L\ a} c_{p-1-L}^{k(p)a} \nonumber \\
&& +
\bar{t}^{(n-p-1)a}_{n-p-2} d * B_{n-p-1\ a}
+ \sum_{L=1}^{n-p-2} \sum_{{k=0 \atop k \ {\rm even}}}^{L} 
\bar{t}^{k(n-p-1)a}_{n-p-2-L} d * t_{n-p-1-L\ a}^{k-1(n-p-1)}
\nonumber \\ && 
+ \sum_{L=1}^{n-p-2} \sum_{{k=1 \atop k \ {\rm odd}}}^{L} 
d * \bar{t}^{k-1(n-p-1)a}_{n-p-1-L} t_{n-p-2-L\ a}^{k(n-p-1)},
\end{eqnarray}
where $*$ is the Hodge star operator.

\subsection{Observables and a generalization of the star product}
\noindent
According to the discussion of the section 3, 
we can take two kind of boundary conditions 
(i) ${\ba_p{}^a}_{//}|_{\partial \Sigma} = 0$ and 
(ii) ${\bb_{n-p-1\ a}}_{//}|_{\partial \Sigma} = 0$
consistently. 
We analyze the total BRST transformations (\ref{totalBRST}) of 
the fields under each boundary condition.
If we take (i), the terms include at least one ${\ba_p{}^a}$.
The total BRST transformation of ${\ba_p{}^a}$ 
at the boundary is as follows:
\begin{eqnarray}
\brs {\ba_p{}^a}_{//}|_{\partial \Sigma}
= d {\ba_p{}^a}_{//}|_{\partial \Sigma}.
\end{eqnarray}
On the other hand, if $S_1$ has no terms which include only one
${\ba_p{}^a}$, $\brs {\bb_{n-p-1\ a}}$ becomes a total derivative:
\begin{eqnarray}
\brs {\bb_{n-p-1\ a}}_{//}|_{\partial \Sigma} 
= d {\bb_{n-p-1\ a}}_{//}|_{\partial \Sigma}, 
\end{eqnarray}
If we take (ii), 
the terms include at least one $\bb_{n-p-1\ a}$.
Then if $S_1$ has no terms which include only one ${\bb_{n-p-1\ a}}$,
the total BRST transformation of ${\ba_p{}^a}$ 
at the boundary becomes 
\begin{eqnarray}
\brs {\ba_p{}^a}_{//}|_{\partial \Sigma}
= d {\ba_p{}^a}_{//}|_{\partial \Sigma}.
\end{eqnarray}
On the other hand, one of ${\bb_{n-p-1\ a}}$ becomes a total
derivative:
\begin{eqnarray}
\brs {\bb_{n-p-1\ a}}_{//}|_{\partial \Sigma} 
= d {\bb_{n-p-1\ a}}_{//}|_{\partial \Sigma}.
\end{eqnarray}

Now we assume that $S_1$ is constructed from terms to satisfy the
condition above. Then we define 
\begin{eqnarray}
&&{\cal O}_{F, p_1, \cdots, p_k, q_1, \cdots, q_l //} \nonumber \\
&& \quad \equiv
\sum_{p_1, \cdots, p_k, q_1, \cdots, q_l}
({F_{p_1 \cdots p_k, q_1 \cdots q_l}
(\ba_0{}^a)
\cdot \ba_{p_1}{}^{a_1} \cdots \ba_{p_k}{}^{a_k}
\cdot \bb_{q_1 b_1} \cdots \bb_{q_l b_l}})_{//}|_{\partial \Sigma}.
\label{obse}
\end{eqnarray}
We can confirm that
\begin{eqnarray}
&& \bDelta {\cal O}_{F, p_1, \cdots, p_k, q_1, \cdots, q_l //} = 0,
\nonumber \\
&& \brs {\cal O}_{F, p_1, \cdots, p_k, q_1, \cdots, q_l //} 
= \sbv{S}{{\cal O}_{F, p_1, \cdots, p_k, q_1, \cdots, q_l //}} 
= d {\cal O}_{F, p_1, \cdots, p_k, q_1, \cdots, q_l //}.
\label{obBRS}
\end{eqnarray}
Especially
\begin{eqnarray}
\bDelta {\cal O}^{(0)}_{F, p_1, \cdots, p_k, q_1, \cdots, q_l //} = 0, 
\qquad \brs {\cal O}^{(0)}_{F, p_1, \cdots, p_k, q_1, \cdots, q_l //}
= 0, 
\end{eqnarray}
where ${\cal O}^{(0)}$ is the $0$-form part of ${\cal O}$.
Therefore ${\cal O}^{(0)}_{F, p_1, \cdots, p_k, q_1, \cdots, q_l}$
is a local observable
with ghost number $|{\cal O}| = p_1 + \cdots + p_k + q_1 + \cdots + q_l$.
We can also construct Wilson-loop-like observables
on the boundary manifold ${\partial \Sigma}$.
\begin{eqnarray}
\int_{\Sigma_r} {\cal O}_{F, p_1, \cdots, p_k, q_1, \cdots, q_l //}
\label{Wilobs}
\end{eqnarray}
where
${\Sigma_r} \subset \partial \Sigma$ is a $r$-dimensional closed
subspace of $\partial \Sigma$ and $0 \leq r \leq |{\cal O}|$.
Actually $\bOmBV \int_{\Sigma_r} {\cal O}_{F, p_1, \cdots, p_k, q_1,
\cdots, q_l //} = 0$ because of the equation (\ref{obBRS}).

We consider simple cases. $\int_{\Sigma_r} {\ba_p{}^a}_{//}$ 
is an observable from (\ref{Wilobs}).
If we take (i), it is trivial.
If we take (ii), it is an observable with ghost number $p-r$.
$\int_{\Sigma_r} {\bb_{n-p-1\ a}}_{//}$
is also an observable from (\ref{Wilobs}).
If we take (i), it is an observable with ghost number $n-p-1-r$.
On the other hand, if we take (ii), it is trivial.
We consider an another example:
\begin{eqnarray}
{\cal O}_{F} \equiv F
(\ba_0{}^a)|_{\partial \Sigma}.
\end{eqnarray}
$0$-form part ${\cal O}_{F}^{(0)}$ of ${\cal O}_{F}$ is an local
observable at the boundary with ghost number zero.

The generating functional is defined as 
\begin{eqnarray}
Z[{\cal O}_k] =
\int \prod_{p=0}^{[\frac{n-1}{2}]} 
{\cal D}{\ba_p{}} {\cal D}{\bb_{n-p-1}} \ 
e^{\frac{i}{\hbar}( S + \sum_r J_r {\cal O}_r)},
\end{eqnarray}
where $J_k$ are source fields and ${\cal O}_k$ are observables.
In two dimension, the correlation function of two local observables 
${\cal O}_f^{(0)}$ and ${\cal O}_g^{(0)}$ leads the star product formula:
\begin{eqnarray}
f * g(x) = \int_{\phi(\infty)=x} 
{\cal D}{\bphi} {\cal D}{\bb_{1}} \ 
{\cal O}^{(0)}_f(\bphi(1)) {\cal O}^{(0)}_g(\bphi(0))
e^{\frac{i}{\hbar} S},
\end{eqnarray}
where $\bphi=\ba_0{}^a$ and $0, 1, \infty$ are 
three distinct points at the boundary $\partial \Sigma$.
We can propose a generalization of the star product to higher
dimensions as follows:
\begin{eqnarray}
m_k[{\cal O}_1, {\cal O}_2, \cdots, {\cal O}_k]
= \int \prod_{p=0}^{[\frac{n-1}{2}]} 
{\cal D}{\ba_p{}} {\cal D}{\bb_{n-p-1}} \ 
{\cal O}_1 {\cal O}_2 \cdots {\cal O}_k
e^{\frac{i}{\hbar}S},
\label{genstar}
\end{eqnarray}
under the appropriate regularization and the boundary conditions,
where $S$ is the deformation (\ref{lapS}) of the $n$-dimensional world
volume BF theory and ${\cal O}_r$'s are observables at the boudanry. 
%
%
%
The correlation functions satisfy the Ward-Takahashi identity 
derived from the gauge symmetry:
\begin{eqnarray}
\int \prod_{p=0}^{[\frac{n-1}{2}]} 
{\cal D}{\ba_p{}} {\cal D}{\bb_{n-p-1}} \ 
\bDelta \left({\cal O} e^{\frac{i}{\hbar}S} \right) = 0,
\end{eqnarray}
where ${\cal O}$ is an observable. The WT identity leads 
$L_{\infty}$ structure on the correlation functions.
Therefore the product (\ref{genstar}) has the $L_{\infty}$ structure.
In two dimension, $L_\infty$-algebra structure on the star
deformation is certainly 
derived from the $L_\infty$-algebra structure of the BRST algebra
\cite{CF}.
We can generalize the structure to higher dimension as 
the WT identity on correlation functions of the higher dimensional BF
theory.

\section{Topological Open Membrane}
\noindent
We consider the theory on the $n-1$-space-dimensional open membrane 
in $N$-dimensional target space.
The open $2$-brane theory with the background $3$-form is analyzed at 
the paper \cite{BBSS}\cite{KS}\cite{Das:2001mg}.

First, we consider the topological membrane theory action
with a background $n$-form field $C$ as follows: 
\begin{eqnarray}
S = \int_{\Sigma} C_{a_1 \cdots a_n}(\phi) d \phi^{a_1} \cdots
d \phi^{a_n},
\label{topmem}
\end{eqnarray}
there $C_{a_1 \cdots a_n}$ is a completely antisymmetric $n$-form.
We rewrite the action (\ref{topmem}) by first order formalism as
follows:
\begin{eqnarray}
S = \int_{\Sigma} B_{n-1 \ a} d \phi^{a}
+ B_{n-1 \ a} A_1{}^{a} + C_{a_1 \cdots a_n} A_1{}^{a_1} \cdots 
A_1{}^{a_n},
\label{topact}
\end{eqnarray}
where $A_1{}^{a}$ and $B_{n-1 \ a}$ 
are auxiliary $1$-form fields and $n-1$-form fields.

In two dimension, the action (\ref{topact}) is written as 
\begin{eqnarray}
S &=& \int_{\Sigma} B_{1 \ a} d \phi^{a}
+ B_{1 \ a} A_1{}^{a} 
+ C_{ab} A_1{}^{a} A_1{}^{b}, \nonumber \\
&=& \int_{\Sigma} B_{1 \ a} d \phi^{a}
- \frac{1}{4} (C^{-1})^{ab} B_{1 \ a} B_{1 \ b} \nonumber \\
&& \qquad \qquad 
+ C_{ab} \left(A_1{}^{a} - \frac{1}{2} (C^{-1})^{ac} B_{1 \ c} \right)
\left(A_1{}^{b} - \frac{1}{2} (C^{-1})^{bd} B_{1 \ d} \right).
\end{eqnarray}
If we set 
$A_1{}^{a\prime} = A_1{}^{a} - \frac{1}{2} (C^{-1})^{ac} B_c$
and integrate out $A_1{}^{a\prime}$, we obtain the action of the
nonlinear gauge theory in two dimension (\ref{twodim}). 

In higher dimensions than two, we can obtain (\ref{topact}) from
our deformed action 
(\ref{lapS}) by the following procedure.
If we take the limit
\begin{eqnarray}
B_{n-p-1\ a} \longrightarrow 0 \quad {\rm for} \quad p \neq 0, 
\nonumber \\
A_{p}{}^a \longrightarrow 0 \quad {\rm for} \quad p \geq 2, 
\end{eqnarray}
we obtain (\ref{topact}), where
$\phi^a = A_0^a$ and
\begin{eqnarray}
F_{1, n-1\ a}{}^b = - \delta_a{}^b,
\qquad
F_{1 \cdots 1, \ a_1 \cdots a_n} = (-1)^n C_{a_1 \cdots a_n}(\phi).
\end{eqnarray}
The topological open membrane is derived as the above scaling
limit of the deformed BF theory.

\section{Conclusion and Discussion}
\noindent
We have considered all possible deformations of the BF theory in any
dimension by the antifield BRST formalism.
We have analyzed the BRST cohomology of the BF theory.
It has led us to a new gauge symmetry and a deformed action.
We have considered quantum BV formalism and quantize
the theory.
We have taken the gauge fixing and found observables.

This gauge symmetry gives an extension to higher dimension 
of the nonlinear gauge symmetry
\cite{II1}\cite{SS} in two dimension, \cite{I2} in three dimension.
We have considered our theory as 
higher dimensional generalization of the two-dimensional nonlinear
gauge theory, or the Poisson sigma model.
It will be useful to analyze topological open membrane 
or noncommutative structure on higher dimensional open membrane. 
In fact, we have 
considered that our action
tend to the topological $n-1$-brane action 
at a certain coupling zero limit.

Our key structure is $L_\infty$-algebra.
Generally, the gauge algebra can be consistently deformed to  
the $L_\infty$-algebra (the strongly homotopy Lie algebra)
in the antifield BRST formalism \cite{LS}\cite{Sta}.
On the other hand, $L_\infty$-algebra is the underline algebra in the 
deformation quantization or formality conjecture \cite{Ko}.
$L_\infty$-algebra structure on the star deformation is certainly
derived from the $L_\infty$-algebra structure of the BRST algebra
at the path integral representation of the deformation quantization
\cite{CF}.
Our extension has respected the similar $L_\infty$-algebra structure.


We can conjecture that 
if we consider the deformation (\ref{lapS}) of the BF theory on the
open membrane, the operator product expansions of the correlation
function at the boundary are deformed and 
(\ref{genstar}) is a deformation of
$n$-algebra \cite{Kontsevich:1999eg}.
The explicit calculations of the correlation
functions are needed on the quantum theory.

In the Schwarz type topological field theory, such as the BF theory,
observables are related to 
miscellaneous topological invariants and knot invariants
\cite{BBRT}\cite{Cattaneo:1995pk}\cite{Cattaneo:1995tw}. 
In our theory, geometrical meanings of correlation 
functions of observables are not still understood. 
More generally, whether deformation of correlation functions 
of a gauge theory is related
to deformation of a mathematical structure or not is not
still known.

Topological field theories called $A$ model and $B$ model 
are introduced to investigate the mirror symmetry
\cite{Witten:1991zz}. 
Different reduction of the topological open membrane theory 
leads us to the $A$ model and $B$ model \cite{Park}.
Our theory may be useful to analyze the mirror symmetry.

\section*{Acknowledgments}
The author thank T.~Asakawa
for discussions and comments about the present work.
\section*{Appendix, Notation}
\noindent
For a superfield $F(\Phi, \Phi^{+})$ and $G(\Phi, \Phi^{+})$,
The following identities are satisfied:
\begin{eqnarray}
&& FG = (-1)^{\gh F \gh G + \deg F \deg G} G F, \nonumber \\
&& d(FG) = dF G + (-1)^{\deg F} F dG, 
\label{FGpro}
\end{eqnarray}
at the usual products.
The graded commutator of two superfields satisfies the following 
identities:
\begin{eqnarray}
&& [F, G] = -(-1)^{\gh F \gh G + \deg F \deg G} [G, F], \nonumber \\
&& [F, [G, H]] = [[F, G], H] 
+ (-1)^{\gh F \gh G + \deg F \deg G} [G, [F, H]].
\label{FGcom}
\end{eqnarray}

We introduce the total degree of a superfield $F$ as 
$|F| = \gh F + \deg F$.
We define the {\it dot product} on superfields as
\begin{eqnarray}
F \cdot G \equiv  (-1)^{\gh F \deg G} FG, 
\label{dotpro}
\end{eqnarray}
and the {\it dot Lie bracket}
\begin{eqnarray}
\lb{F}{G} \equiv (-1)^{\gh F \deg G} [F, G].
\label{dotlie}
\end{eqnarray}
We obtain the following identities
of the dot product and the dot Lie bracket from
(\ref{FGpro}), (\ref{FGcom}), (\ref{dotpro}) and (\ref{dotlie}):
\begin{eqnarray}
&& F \cdot G = (-1)^{|F||G|} G \cdot F, \nonumber \\
&& \lb{F}{G} = - (-1)^{|F||G|} \lb{G}{F}, \nonumber \\
&& \lb{F}{\lb{G}{H}} = \lb{\lb{F}{G}}{H}
+ (-1)^{|F||G|} \lb{G}{\lb{F}{H}},
\end{eqnarray}
and
\begin{eqnarray}
d (F \cdot G) \equiv d F \cdot G + (-1)^{|F|} F \cdot d G.
\end{eqnarray}
\newcommand{\bibit}{\sl}


\vfill\eject
\end{document}